\documentclass[a4paper,12pt]{article}
\usepackage{amsmath}
\usepackage{accents}
\usepackage[english]{babel}
\usepackage{graphicx}
\usepackage[usenames]{color}
\usepackage{colortbl}

\begin{document}

\centerline {\LARGE{Geometry of quantum state manifolds generated}}
\centerline {\LARGE{by the Lie algebra operators}}
\medskip
\centerline {A. R. Kuzmak}
\centerline {\small \it E-Mail: $^1$andrijkuzmak@gmail.com}
    \medskip
\centerline {\small \it Department for Theoretical Physics, Ivan Franko National University of Lviv,}
\medskip
\centerline {\small \it 12 Drahomanov St., Lviv, UA-79005, Ukraine}

{\small

The Fubini-Study metric of quantum state manifold generated by the operators which satisfy the Heisenberg Lie algebra is calculated.
The similar problem is studied for the manifold generated by the $so(3)$ Lie algebra operators.
Using these results, we calculate the Fubini-Study metrics of state manifolds generated by the position and momentum operators.
Also the metrics of quantum state manifolds generated by some spin systems are obtained. Finally, we generalize this problem for operators of
an arbitrary Lie algebra.

\medskip

}

\section{Introduction\label{sec1}}

The investigation of the geometrical properties of quantum state manifolds is often useful in the study of the evolution of quantum systems \cite{gqev1,gqev2,gqev3,FSM2}.
The information about the geometry of manifold which contains all states achieved during the evolution of quantum system
allows us to simplify the exploration of the system dynamics. For instance, the Hamiltonian which provides the time-optimal evolution between
two quantum states was obtained using the symmetry properties of quantum state space \cite{OHfST}. This is the solution of
the so-called quantum brachistochrone problem \cite{brach1}. The conditions for time-optimal evolution of a spin-$\frac{1}{2}$ in the magnetic field
were obtained using the fact that the whole state space of this system is represented by the Bloch sphere \cite{FSM2,TMTSPMF,FSM,TOSTSS}.
So, it was obtained that the optimal evolution happens when the magnetic field is orthogonal to the initial and the final states.
Then the state evolves through a trajectory which is a geodesic line on the sphere. Similar problems for the spin-$1$ and arbitrary spin
were solved in \cite{QBS1} and \cite{brachass}, respectively. Also the geometry properties of multilevel quantum systems are intensively examined
\cite{mlqsg0,mlqsg1,mlqsg2,mlqsg3,mlqsg4,mlqsg5,mlqsg6, mlqsg7} because they could be more efficient for quantum computation
than qubits \cite{qcomp6,qcomp7,qcomp8}. It was shown that the problem of finding of the quantum circuit of unitary operators
which provide time-optimal evolution on a system of qubits \cite{OCGQC,GAQCLB,QCAG,QGDM} and qutrits \cite{GQCQ} is related to the problem of
finding the minimal distance between two point on the Riemannian metric. These results are important for the implementation
of quantum computations \cite{qcomp1,qcomp2,qcomp3,qcomp4}. For example, in \cite{qcomp5} it was shown how the geometry
of $n$-qubit quantum space is associated with realization of quantum computations.

The method, which allows us to study the geometry of quantum state manifolds, is based on the investigation of their Fubini-Study metric.
The Fubini-Study metric is defined by the infinitesimal distance $ds$ between two neighboring pure quantum states
$\vert\psi (\xi^{\mu})\rangle$ and $\vert\psi (\xi^{\mu}+d\xi^{\mu})\rangle$
\begin{eqnarray}
ds^2=g_{\mu\nu}d\xi^{\mu}d\xi^{\nu},
\label{form5}
\end{eqnarray}
where $\xi^{\mu}$ is a set of real parameters which define the state $\vert\psi(\xi^{\mu})\rangle$. The components of the metric tensor
$g_{\mu\nu}$ have the form
\begin{eqnarray}
g_{\mu\nu}=\gamma^2\Re\left(\langle\psi_{\mu}\vert\psi_{\nu}\rangle-\langle\psi_{\mu}\vert\psi\rangle\langle\psi\vert\psi_{\nu}\rangle\right),
\label{form6}
\end{eqnarray}
where $\gamma$ is an arbitrary factor which is often chosen to have value of $1$, $\sqrt{2}$ or $2$ and
\begin{eqnarray}
\vert\psi_{\mu}\rangle=\frac{\partial}{\partial\xi^{\mu}}\vert\psi\rangle.
\label{form7}
\end{eqnarray}
One can find more about this form of metric in papers \cite{gqev2,FSM2,FSM,FSM3,FSM0,FSM1,FSMqsgm,FSMqsgm2}.

In Refs. \cite{FSMqsgm,CSRS} the Fubini-Study metrics of some well known coherent state manifolds were obtained.
In the articles the authors considered the atomic coherent states, generated by the action of the $SU(2)$ displacement operator
on the eigenstate of the $z$-component of the angular momentum operator which correspond to the lowest eigenvalue.
They obtained that the metric of manifold defined by these states is that of the sphere. In \cite{brachass} this problem was generalized
in the case of the manifold which contains the states achieved by the rotation of the eigenstate of the operator of projection
of arbitrary spin on some direction. In this case the manifold is the sphere with the radius dependent on the value of the spin
and on the value of the spin projection. In another paper \cite{FSM4} the geometry of the ground state manifold of the quantum $XY$
chain in a transverse magnetic field was considered. In this case the authors found that the metric tensor of this manifold
depends on the exchange coupling and on the value of magnetic field. In our previous papers \cite{torus,FMM} we studied
the quantum evolution of a two-spin systems with isotropic and anisotropic interactions in the magnetic field.
The interaction and the magnetic-field parts of the Hamiltonian commute between themselves. As a result we obtained that the metric
of quantum state manifold is flat and depends on the ratio between interaction couplings and on the parameters of the initial states.
In the present paper, we investigate the Fubini-Study metrics of manifolds generated by the Heisenberg Lie algebra (Section \ref{sec2})
and $so(3)$ Lie algebra (Section \ref{sec3}) operators. Thus, we calculate the metrics of the quantum state manifolds
of some physical implementations. Finally, we obtain the Fubini-Study metric of quantum state manifold generated
by an arbitrary Lie algebra (Section \ref{sec4}).

\section{The Fubini-Study metric of quantum states manifold generated by the Heisenberg Lie algebra operators \label{sec2}}

We consider the following unitary operator
\begin{eqnarray}
U=\exp\left(-i\sum_{j=1}^n\theta_j A_j\right)\exp\left(-i\sum_{j=1}^n\phi_j B_j\right),
\label{form1}
\end{eqnarray}
where $\theta_j$, $\phi_j$ are some real parameters and $A_j$, $B_j$ are Hermitian operators, which satisfy the following Lie brackets
\begin{eqnarray}
\left[A_j,B_l\right]=i\delta_{jl}C,\quad \left[A_j,C\right]=\left[B_j,C\right]=0,
\label{form2}
\end{eqnarray}
where $\delta_{jl}$ is the Kronecker delta. These brackets generate the Heisenberg Lie algebra. Also $\left[A_j,A_l\right]=\left[B_j,B_l\right]=0$.

The unitary transformation of quantum state $\vert\psi_i\rangle$ under the action of operator (\ref{form1}) can be represented as follows
\begin{eqnarray}
\vert\psi\rangle=U\vert\psi_i\rangle.
\label{form4}
\end{eqnarray}
This state defined by real parameters $\theta_j$ and $\phi_j$. Let us calculate the Fubini-Study metric of the manifold defined
by these parameters. First of all, we calculate the following scalar products
\begin{eqnarray}
&&\langle\psi\vert\psi_{\theta_j}\rangle = -i\langle \tilde{A_j}\rangle, \quad   \langle\psi_{\theta_j}\vert\psi_{\theta_l}\rangle = \langle \tilde{A_j}\tilde{A_l}\rangle,\nonumber\\
&&\langle\psi\vert\psi_{\phi_j}\rangle = -i\langle B_j\rangle,\quad \langle\psi_{\phi_j}\vert\psi_{\phi_l}\rangle = \langle B_jB_l\rangle,\nonumber\\
&&\langle\psi_{\theta_j}\vert\psi_{\phi_l}\rangle = \langle \tilde{A_j}B_l\rangle,
\label{form8}
\end{eqnarray}
where a bracket $\langle A\rangle$ denotes the expectation value with respect to $\vert\psi_i\rangle$. Here we introduce the following
notation ${\tilde A_j}=A_j+\phi_j C$.
Substituting these products into (\ref{form6}), we obtain the following components of the metric tensor
\begin{eqnarray}
g_{\theta_j\theta_l}=\gamma^2\langle\Delta {\tilde A_j}\Delta {\tilde A_l}\rangle,\quad g_{\phi_j\phi_l}=\gamma^2\langle\Delta B_j \Delta B_l\rangle,\quad g_{\theta_j\phi_l}=\frac{\gamma^2}{2}\langle \{\Delta {\tilde A_j},\Delta B_l\}\rangle ,
\label{form9}
\end{eqnarray}
where $\Delta A= A - \langle A\rangle$ is the standard deviation operator and
$\left\{A,B\right\}=AB+BA$ is the anticommutator of two operators.
The metric (\ref{form9}) is defined by the $2n$ local basis vectors
\begin{eqnarray}
\vert\psi_j\rangle=\gamma\Delta \tilde{A_j}\vert\psi_i\rangle, \quad \vert\psi_j\rangle=\gamma\Delta B_j\vert\psi_i\rangle.
\label{form9_1}
\end{eqnarray}
It is easy to see that the components of the metric tensor do not depend on the parameters $\theta_j$ and $\phi_j$ when the commutator
between operator $A_j$ and $B_j$ equals some constant or zero. Then we have the metric of the flat manifold \cite{FMM}
\begin{eqnarray}
g_{\theta_j\theta_l}=\gamma^2\langle\Delta A_j\Delta A_l\rangle,\quad g_{\phi_j\phi_l}=\gamma^2\langle\Delta B_j\Delta B_l\rangle,\quad g_{\theta_j\phi_l}=\frac{\gamma^2}{2}\langle \{\Delta A_j,\Delta B_l\}\rangle .
\label{form10}
\end{eqnarray}

For {\it example}, let us consider the case when the Heisenberg Lie algebra is generated by the position ${\bf r}$, momentum ${\bf p}$ and unit $I$ operators
\begin{eqnarray}
\left[x_j,p_l\right]=i\delta_{jl},\quad \left[x_j,I\right]=\left[p_j,I\right]=0,
\label{form11}
\end{eqnarray}
where $x_j$ and $p_j$ are the components of the position and momentum operators, respectively.
We set $\hbar =1$. Then the state generated by the action of the operator (\ref{form1}) with $A_j=x_j$, $B_j=p_j$ on the initial state
$\vert\psi_i\rangle$ can be expressed as follows
\begin{eqnarray}
\vert\psi\rangle=\exp\left(-i\sum_{j=1}^3\theta_j x_j\right)\exp\left(-i\sum_{j=1}^3\phi_j p_j\right)\vert\psi_i\rangle.
\label{form12}
\end{eqnarray}
In this equation the first and second multipliers are translation operators in position and momentum spaces, respectively.
Since the operator $C=1$ is a constant, the metric of the manifold defined by transformation (\ref{form12}) is calculated using equations (\ref{form10}).
So, the metric of the manifold generated by algebra (\ref{form11}) is flat,
\begin{eqnarray}
g_{\theta_j\theta_l}=\gamma^2\langle\Delta x_j\Delta x_l\rangle,\quad g_{\phi_j\phi_l}=\gamma^2\langle\Delta p_j\Delta p_l\rangle,\quad g_{\theta_j\phi_l}=\frac{\gamma^2}{2}\langle \{\Delta x_j,\Delta p_l\}\rangle .
\label{form10_1}
\end{eqnarray}
For instance, let us take the initial state $\vert\psi_i\rangle$ of the one-dimensional harmonic oscillator $\vert n\rangle$.
Then, the momentum $x$ and position $p$ operators can be represented by the bosonic creation $a^+$ and annihilation $a$ operators
\begin{eqnarray}
x=\frac{1}{\sqrt{2m\omega}}(a^++a),\quad p=i\sqrt{\frac{m\omega}{2}}(a^+-a),
\label{form12_2}
\end{eqnarray}
where $m$ and $\omega$ are the mass and the angular frequency of the oscillator, respectively. Using equations (\ref{form10_1})
with expressions (\ref{form12_2})  we obtain
\begin{eqnarray}
g_{\theta\theta}=\frac{\gamma^2}{2m\omega}(2n+1),\quad g_{\phi\phi}=-\frac{\gamma^2m\omega}{2}(2n+1),\quad g_{\theta\phi}=0.
\label{form12_3}
\end{eqnarray}
This is the metric of the infinite plane which contains all states obtained by transformation (\ref{form12}) with
$\theta\in[-\infty,+\infty]$ and $\phi\in[-\infty,+\infty]$.

\section{The Fubini-Study metric of quantum states manifold generated by $so(3)$ Lie algebra operators  \label{sec3}}

In the present section, we study the Fubini-Study metric of the manifold which contains the quantum states obtained by the action
of the $SO(3)$ group element $U$ on the initial state $\vert\psi_i\rangle$. It is well known that an arbitrary element of $SO(3)$
group can be expressed using the elements from one-parametric subgroup as follows
\begin{eqnarray}
U=e^{-i\theta_1A_1}e^{-i\theta_2A_2}e^{-i\theta_3A_1},
\label{form13}
\end{eqnarray}
where $\theta_i$ are the Euler angles and $A_i$ are the Hermitian operators which satisfy the commutation relations
\begin{eqnarray}
[A_1,A_2]=iA_3,\quad [A_2,A_3]=iA_1, \quad [A_3,A_1]=iA_2.
\label{form14}
\end{eqnarray}
Similarly to the previous case, we calculate the Fubini-Study metric of the manifold generated by transformation (\ref{form4})
with unitary operator (\ref{form13}). So, we obtain the components of the metric tensor
\begin{eqnarray}
&&g_{\theta_1\theta_1}=\gamma^2\langle\Delta \tilde{A}_3^2\rangle,\quad g_{\theta_2\theta_2}=\gamma^2\langle\Delta \tilde{A}_2^2\rangle,\quad g_{\theta_3\theta_3}=\gamma^2\langle\Delta A_1^2\rangle,\nonumber\\
&&g_{\theta_1\theta_2}=\frac{\gamma^2}{2}\langle\{\Delta \tilde{A}_2,\Delta \tilde{A}_3\}\rangle,\quad g_{\theta_1\theta_3}=\frac{\gamma^2}{2}\langle\{\Delta A_1,\Delta \tilde{A}_3\}\rangle,\nonumber\\
&&g_{\theta_2\theta_3}=\frac{\gamma^2}{2}\langle\{\Delta A_1,\Delta \tilde{A}_2\}\rangle,
\label{form15}
\end{eqnarray}
where we introduce the following notations
\begin{eqnarray}
\tilde{A}_2=\cos\theta_3 A_2-\sin\theta_3 A_3,\quad \tilde{A}_3=\cos\theta_3 A_1+\sin\theta_2\sin\theta_3 A_2+\sin\theta_2\cos\theta_3 A_3.\nonumber
\end{eqnarray}
Expressions (\ref{form15}) describe the metric of the three-parametric quantum states manifold defined by the three local basis vectors
\begin{eqnarray}
\vert\psi_1\rangle=\gamma\Delta A_1\vert\psi_i\rangle,\quad \vert\psi_2\rangle=\gamma\Delta \tilde{A}_2\vert\psi_i\rangle, \quad \vert\psi_3\rangle=\gamma\Delta \tilde{A}_3\vert\psi_i\rangle.
\label{form16}
\end{eqnarray}

Let us consider some physical examples.

{\it Example I}. In this example we consider the rotation of the quantum state of spin-$s$. The components
of the spin-$s$ operator ${\bf S}=(S_x,S_y,S_z)$ are the generator of $so(3)$ algebra (\ref{form14}) with
$A_1\equiv S_z$, $A_2\equiv S_x$, $A_3\equiv S_y$. The arbitrary rotation of quantum state of spin $\vert\psi_i\rangle$
can be realized using (\ref{form13}). It can be implemented if the magnetic field is applied along the respective directions.
Then the role of the Euler angles plays $\theta_1=ht_1$, $\theta_2=ht_2$ and $\theta_3=ht_3$, where $h$ is the value
of the magnetic field and $t_i$ is the period of the action of the magnetic field along specific direction.
So, the metric of quantum manifold defined by these rotations is described by expression (\ref{form15}). If the initial state is
the eigenstate of the $S_z$ operator with the eigenvalue $m$ then
\begin{eqnarray}
&&\langle\Delta S_x^2\rangle =\frac{1}{2}\left(s(s+1)-m^2\right),\quad \langle\Delta S_y^2\rangle =\frac{1}{2}\left(s(s+1)-m^2\right),\nonumber\\
&&\langle\Delta S_z^2\rangle =0,\quad \langle\{\Delta S_i,\Delta S_j\}\rangle=0\quad i\neq j\nonumber
\end{eqnarray}
and metric takes the form \cite{brachass}
\begin{eqnarray}
ds^2=R^2\left[\sin^2\theta_2\right(d\theta_1\left)^2+\right(d\theta_2\left)^2\right].
\label{form19}
\end{eqnarray}
where $R=\left(\gamma/\sqrt{2}\right)\sqrt{s(s+1)-m^2}$. This is the metric of the sphere of radius $R$.
As we can see this manifold is two-parametric. Thus, the evolution of spin-$s$ between two states on this sphere is
provided by the rotation of the spin about some axis defined by unit vector \cite{brachass}.

In the general case when the initials state is a superposition of the eigenstates $\vert m\rangle$
\begin{eqnarray}
\vert\psi_i\rangle=\sum_{m=-s}^sC_m\vert m\rangle
\label{form19_1}
\end{eqnarray}
the geometry of state manifold is more difficult and is defined by three parameters. Here $C_m$ is a set complex parameters
which define the initial state and satisfy the normalization condition $\vert\psi_i\rangle=\sum_{m=-s}^s\vert C_m\vert^2$.
For instance, if we consider the following state $\vert\psi_i\rangle=C_{-1}\vert -1\rangle+C_{1}\vert 1\rangle$
of spin-$1$ system then the components of metric tensor takes the form
\begin{eqnarray}
&&g_{\theta_1\theta_1}=\gamma^2[\cos^2\theta_2\langle\Delta S_z^2\rangle+\sin^2\theta_2\sin^2\theta_3\langle\Delta S_x^2\rangle
+\sin^2\theta_2\cos^2\theta_3\langle\Delta S_y^2\rangle\nonumber\\
&&+\sin^2\theta_2\cos\theta_3\sin\theta_3\langle\left\{\Delta S_x,\Delta S_y\right\}\rangle],\nonumber\\
&&g_{\theta_2\theta_2}=\gamma^2[\cos^2\theta_3\langle\Delta S_x^2\rangle+\sin^2\theta_3\langle\Delta S_y^2\rangle-\cos\theta_3\sin\theta_3\langle\left\{\Delta S_x,\Delta S_y\right\}\rangle],\nonumber\\
&&g_{\theta_3\theta_3}=\gamma^2\langle\Delta S_z^2\rangle,\nonumber\\
&&g_{\theta_1\theta_2}=\frac{\gamma^2}{2}[\sin\theta_2\sin 2\theta_3\left(\langle\Delta S_x^2\rangle-\langle\Delta S_y^2\rangle\right)
+\sin\theta_2\cos 2\theta_3\langle\left\{\Delta S_x,\Delta S_y\right\}\rangle],\nonumber\\
&&g_{\theta_1\theta_3}=\gamma^2\cos\theta_2\langle\Delta S_z^2\rangle,\nonumber\\
&&g_{\theta_2\theta_3}=0,
\label{form19_2}
\end{eqnarray}
where
\begin{eqnarray}
&&\langle\Delta S_z^2\rangle=1-(1-2\vert C_{-1}\vert^2),\quad \langle\Delta S_x^2\rangle=-\langle\Delta S_y^2\rangle=\frac{1}{\sqrt{2}}\Re(C_1^*C_{-1})+\frac{1}{2},\nonumber\\
&&\left\{\Delta S_x,\Delta S_y\right\}=\sqrt{2}\Im(C_1^*C_{-1}).\nonumber
\end{eqnarray}

{\it Example II}. Let us consider the two spin-$\frac{1}{2}$ system which are described by the Hamiltonian
\begin{eqnarray}
H=J_1\left(S_x^1S_y^2-S_y^1S_x^2\right)+J_2\left(S_x^1S_x^2+S_y^1S_y^2\right)+\frac{h_z}{2}\left(S_z^1-S_z^2\right),
\label{form20}
\end{eqnarray}
where $S_i^1=S_i\otimes 1$, $S_i^2=1\otimes S_i$, $S_i$ ($i=x,y,z$) are the spin components, $J_k$ ($k=1,2$) are the interaction coupling,
and $h_z$ is proportional to the value of the magnetic field. The first term is the $z$-component of the Dzyaloshinsky-Moria Hamiltonian,
the second term describes the $XX$ interaction between spins and the third term describes the interaction of the magnetic field with
spins. It is important to note that the effective spins model can be prepared on ultracold atoms in optical lattice
\cite{opticallattice1,opticallattice18}. So, using the methods based on the interference of laser beams \cite{opticallattice17}
we can provide individually to each spins the effective magnetic field. This allows to apply to the spins the opposite magnetic fields
along the $z$-axis.

We can see that the operators in Hamiltonian (\ref{form20}) are the generators of $so(3)$ algebra (\ref{form14})
if we introduce the notation
\begin{eqnarray}
A_1=\frac{1}{2}\left(S_z^1-S_z^2\right),\quad A_2=S_x^1S_y^2-S_y^1S_x^2,\quad A_3=S_x^1S_x^2+S_y^1S_y^2.
\label{form21}
\end{eqnarray}
Here the operator of evolution can be also expressed in the form (\ref{form13})
with Euler angles $\theta_1=h_zt_1$, $\theta_2=J_1t_2$ and $\theta_3=h_zt_3$. It is easy to find the relations of the
time of evolution in $U=\exp{\left(-iHt\right)}$ and the Euler angles. For this purpose, we equate the operator of evolution
with Hamiltonian (\ref{form20}) to unitary operator (\ref{form13}) with (\ref{form21}). Then we obtain
\begin{eqnarray}
\tan\frac{\theta_1+\theta_3}{2}=\frac{h_z}{2\omega}\tan(\omega t),\quad \tan\frac{\theta_1-\theta_3}{2}=\frac{J_2}{J_1},\quad
\cos\frac{\theta_2}{2}\cos\frac{\theta_1+\theta_3}{2}=cos(\omega t).
\label{relateuler}
\end{eqnarray}

It is worth noting that Hamiltonian (\ref{form20}) provides the evolution on subspace spanned by $\vert\uparrow\downarrow\rangle$
and $\vert\downarrow\uparrow\rangle$ vectors \cite{brach}. Let us put the initial state in the form $\vert\psi_i\rangle=\vert\uparrow\downarrow\rangle$ or $\vert\downarrow\uparrow\rangle$.
Then we obtain that in this case the evolution of system happens on manifold with metric of sphere (\ref{form19})
of radius $\gamma/2$. In this case similarly as in the previous example the evolution is determined by two parameters.
However if the initial state has the form $\vert\psi_i\rangle=a\vert\uparrow\downarrow\rangle+b\vert\downarrow\uparrow\rangle$
then the evolution of the system happens on the three-parametric manifold.

Also, there exists a similar Hamiltonian of two spins which provides the quantum evolution on the subspace spanned by
$\vert\uparrow\uparrow\rangle$ and $\vert\downarrow\downarrow\rangle$ vectors and consists of the operators which satisfy the Lie
algebra (\ref{form14})
\begin{eqnarray}
H=J_1\left(S_x^1S_y^2+S_y^1S_x^2\right)+J_2\left(S_x^1S_x^2-S_y^1S_y^2\right)+\frac{h_z}{2}\left(S_z^1+S_z^2\right).
\label{form23}
\end{eqnarray}
In this case the generators of $so(3)$ algebra (\ref{form14}) are the following
\begin{eqnarray}
A_1=\frac{1}{2}\left(S_z^1+S_z^2\right),\quad A_2=S_x^1S_x^2-S_y^1S_y^2,\quad A_3=S_x^1S_y^2+S_y^1S_x^2.
\label{form24}
\end{eqnarray}
Similarly as in the previous case we can generate the evolution of two-spin system using the unitary operator (\ref{form13}) with
Euler angles $\theta_1=h_zt_1$, $\theta_2=J_2t_2$ and $\theta_3=h_zt_3$. We obtain that if the initial state is
$\vert\uparrow\uparrow\rangle$ or $\vert\downarrow\downarrow\rangle$ then evolution of system happens
on manifold which is the sphere of radius $\gamma/2$.

{\it Example III}. The previous example can be generalized as follows
\begin{eqnarray}
H=J_1A_1+J_2A_2+hA_3,
\label{form25}
\end{eqnarray}
where
\begin{eqnarray}
A_1=\frac{{\bf n}}{2}\cdot\left({\bf S}^1-{\bf S}^2\right),\quad A_2={\bf n} \cdot {\bf S}^1\times {\bf S}^2,\quad A_3={\bf S}^1\cdot {\bf S}^2-{\bf n}\cdot{\bf S}^1{\bf n}\cdot{\bf S}^2,
\label{form26}
\end{eqnarray}
where ${\bf n}=(\sin\eta\cos\chi,\sin\eta\sin\chi,\cos\eta)$ is the unit vector represented by the polar $\eta$ and azimuthal $\chi$ angles.
These operators satisfy the Lie algebra (\ref{form14}). The first term in Hamiltonian (\ref{form25}) defines the
direction of the magnetic field, the second term describes the Dzyaloshinsky-Moria interaction, and the third term is the Heisenberg Hamiltonian.
Hamiltonian (\ref{form26}) provides the evolution on subspace spanned by $\vert\!+-\!\rangle$ and
$\vert\!-+\!\rangle$, where $\vert\!+\!\rangle\!=\!\cos\left(\eta/2\right)\vert\!\uparrow\!\rangle +\sin\left(\eta/2\right)e^{i\chi}\vert\uparrow\rangle$ and
$\vert -\rangle=-\sin\left(\eta/2\right)\vert\uparrow\rangle+\cos\left(\eta/2\right)e^{i\chi}\vert\uparrow\rangle$ are the states of spin-$1/2$ projected on
the positive and negative directions of the unit vector ${\bf n}$, respectively. So, in general case the evolution of two-spin system
happens on the three-parametric manifold with metric defined by expression (\ref{form15}). However, if the initial state is
$\vert+-\rangle$ or $\vert-+\rangle$ then the manifold is two-parametric sphere of radius $\gamma/2$.

\section{Generalization for an arbitrary Lie algebra \label{sec4}}

The problem which we considered in the previous sections can be generalized for an arbitrary Lie algebra defined by $N$ operators $A_i$,
which satisfy the following brackets
\begin{eqnarray}
\left[A_i,A_j\right]=\sum_{k=1}^Nc_{ij}^kA_k,
\label{form27}
\end{eqnarray}
where $c_{ij}^k$ are the structure constants which satisfy the following condition $c_{ij}^k=-c_{ji}^k$. Let us study
Fubini-Study metric of the manifold determined by unitary transformation (\ref{form4}) with the following operator
\begin{eqnarray}
U=e^{-i\theta_1A_1}e^{-i\theta_2A_2}\ldots e^{-i\theta_NA_N},
\label{form28}
\end{eqnarray}
where $\theta_i$ are some real parameters. Similarly as in the previous cases,
we calculate the components of the metric tensor
\begin{eqnarray}
g_{\theta_i\theta_j}=\frac{\gamma^2}{2}\langle\{\Delta \tilde{A}_i,\Delta \tilde{A}_j\}\rangle.
\label{form29}
\end{eqnarray}
The operators $\tilde{A}_j$ are obtained using the Baker-Campbell-Hausdorff formula with the Lie brackets (\ref{form27})
\begin{eqnarray}
&&\tilde{A}_j=e^{i\theta_NA_N}e^{i\theta_{N-1}A_{N-1}}\ldots e^{i\theta_{j+1}A_{j+1}}A_je^{-i\theta_{j+1}A_{j+1}}\ldots e^{-i\theta_{N-1}A_{N-1}}e^{-i\theta_NA_N}\nonumber\\
&&=A_j+\sum_{l=1}^{N-j}\sum_{n_l=l}^{N-j}\ldots\sum_{n_2=2}^{n_3-1}\sum_{n_1=1}^{n_2-1}C_{j+n_1\ j}C_{j+n_1}^{-1}(e^{i\theta_{j+n_1}C_{j+n_1}}-1)\nonumber\\
&&\times(e^{i\theta_{j+n_2}C_{j+n_2}}-1)\ldots (e^{i\theta_{j+n_l}C_{j+n_l}}-1)A,
\label{form30}
\end{eqnarray}
where
\begin{eqnarray}
&&C_{j+n_1\ j}=\left(c_{j+n_1\ j}^1,\ c_{j+n_1\ j}^2,\ \ldots\ c_{j+n_1\ j}^N\right),\nonumber\\
&&C_{j+n_l}=\left( \begin{array}{ccccc}
c_{j+n_l\ 1}^1 & c_{j+n_l\ 1}^2 & \ldots & c_{j+n_l\ 1}^N\\
c_{j+n_l\ 2}^1 & c_{j+n_l\ 2}^2 & \ldots & c_{j+n_l\ 2}^N \\
\ldots & \ldots & \ldots & \ldots \\
c_{j+n_l\ N}^1 & c_{j+n_l\ N}^2 & \ldots & c_{j+n_l\ N}^N
\end{array}\right),\quad
A=\left( \begin{array}{ccccc}
A_1\\
A_2\\
\ldots\\
A_N
\end{array}\right).\nonumber
\end{eqnarray}

\section{Conclusion \label{sec6}}

We studied the geometry of the manifold which contains all quantum states achieved during
the unitary transformation generated by $2n$ operators which in turn satisfy Heisenberg Lie algebra (\ref{form2}).
As a result the Fubini-Study metric of manifold was obtained. This metric is defined by $2n$ local
basis vectors (\ref{form9_1}) which describe the flat manifold when the commutators between the operators equal some constant
or zero and the curved manifold in other cases. For instance, we considered the case
when the state manifold is generated by the momentum and position operators. We obtained that it is an infinite space
where each point is determined by the translations of the initial state in momentum and position spaces,
respectively. Also we solved the similar problem when the quantum state manifold is generated by the operators which satisfy
$so(3)$ Lie algebra (\ref{form14}). We calculated metric tensors of manifold (\ref{form15}) and obtained
local basis vectors (\ref{form16}) which define it. Using these results we studied the geometry of quantum state manifolds of some
physical systems, namely, the arbitrary spin in the magnetic field and two-spin system described by the Hamiltonian with Heisenberg and
Dzyaloshinsky-Moria interactions in the magnetic field. Finally, we calculated the Fubini-Study metric of quantum state manifold
generated by arbitrary Lie algebra operators (\ref{form29}).

\section{Acknowledgements}

First of all, the author thanks Prof. Volodymyr Tkachuk for his great support during the study of the problem.
The author thanks Drs. Volodymyr Pastukhov, Askold Duviryak and Andrij Rovenchak for useful comments. This work was supported
by Project FF-30F (No. 0116U001539) from the Ministry of Education and Science of Ukraine.


\begin{thebibliography}{99}
\bibitem{gqev1} J. Anandan, and Y. Aharonov, Phys. Rev. Lett. \textbf{65}, 1697 (1990).
\bibitem{gqev2} S. Abe, Phys. Rev. A {\bf 48}, 4102 (1993).
\bibitem{gqev3} H. P. Laba, V. M. Tkachuk, Cond. Matt. Phys. {\bf 20}, 13003 (2017).
\bibitem{FSM2} I. Bengtsson and K. \.Zyczkowski, {\it Geometry of quantum states}, (New York: Cambridge University press, 2006).
\bibitem{OHfST} D. C. Brody, and D. W. Hook, J. Phys. A \textbf{39}, L167 (2006).
\bibitem{brach1} A. Carlini, A. Hosoya, T. Koike, Y. Okudaira, Phys. Rev. Lett. {\bf 96}, 060503 (2006).
\bibitem{TMTSPMF} U. Boscain, P. Mason, J. Math. Phys. {\bf 47}, 062101 (2006).
\bibitem{FSM} V. M. Tkachuk, {\it Fundamental problems of quantum mechanic} (Lviv: Ivan Franko National University of Lviv, 2011). [in Ukrainian]
\bibitem{TOSTSS} A. D. Boozer, Phys. Rev. A \textbf{85}, 012317 (2012).
\bibitem{QBS1} A. M. Frydryszak and V. M. Tkachuk, Phys. Rev. A {\bf 77}, 014103 (2008).
\bibitem{brachass} A. R. Kuzmak, V. M. Tkachuk, Phys. Lett. A \textbf{379}, 1233 (2015).
\bibitem{mlqsg0} L. Jacobczyk and M. Siennicki, Phys. Lett. A {\bf 286}, 383 (2001).
\bibitem{mlqsg1} D. C. Brody and L. P. Hughston, Journal of Geometry and Physics {\bf 38}, 19 (2001).
\bibitem{mlqsg2} G. Kimura, Phys. Lett. A {\bf 314}, 339 (2003).
\bibitem{mlqsg3} G. Kimura, A. Kossakowski, Open Sys. Inf. Dyn. {\bf 12}, 207 (2005).
\bibitem{mlqsg4} I. P. Mendas, J. Phys. A {\bf 39}, 11313 (2006).
\bibitem{mlqsg5} R. A. Bertlmann and P. Krammer, J. Phys. A {\bf 41}, 235303 (2008).
\bibitem{mlqsg6} S. K. Goyal, B. Neethi Simon, R. Singh and S. Simon, J. Phys. A {\bf 49}, 165203 (2016).
\bibitem{mlqsg7} P. Kurzynski, A. Kolodziejski, W. Laskowski and M. Markiewicz, Phys. Rev. A {\bf 93}, 062126 (2016).
\bibitem{qcomp6} M. Hillery, V. Buzek and M. Ziman, Phys. Rev. A {\bf 65}, 022301 (2002).
\bibitem{qcomp7} D. L. Zhou, B. Zeng, Z. Xu, and C. P. Sun, Phys. Rev. A {\bf 68}, 062303 (2003).
\bibitem{qcomp8} S. S. Bullock, D. P. O’Leary, and G. K. Brennen, Phys. Rev. Lett. {\bf 94}, 230502 (2005).
\bibitem{OCGQC} M. A. Nielsen, M. R. Dowling, M. Gu and A. C. Doherty, Phys. Rev. A {\bf 73}, 062323 (2006).
\bibitem{GAQCLB} M. A. Nielsen, Quant. Inform. Comput. {\bf 6}, 213 (2006).
\bibitem{QCAG} M. A. Nielsen, M. R. Dowling, M. Gu and A. C. Doherty, Science {\bf 311}, 1133 (2006).
\bibitem{QGDM} N. Khaneja, B. Heitmann, A. Sp\"orl, H. Yuan, T. Schulte-Herbr\"uggen and S. J. Glaser, arXiv:quant-ph/0605071 (2006).
\bibitem{GQCQ} Bin Li, Zu-Huan Yu and Shao-Ming Fei, Nature Scientific Report {\bf 3}, 2594 (2013).
\bibitem{qcomp1} A. Barenco, C. H. Bennett, R. Cleve, D. P. DiVincenzo, N. Margolus, P. Shor, T. Sleator, J. A. Smolin, H. Weinfurter, Phys. Rev. A. {\bf 52} 3457 (1995).
\bibitem{qcomp2} B. E. Kane, Nature {\bf 393}. 133 (1998).
\bibitem{qcomp3} D. P. DiVincenzo, D. Bacon, J. Kempe, G. Burkard and K. B. Whaley, Nature {\bf 408}, 339 (2000).
\bibitem{qcomp4} J. Clarke and F. K. Wilhelm, Nature {\bf 453}, 1031 (2008).
\bibitem{qcomp5} A. J. Hanson. G. Ortiz, A. Sabry and Yu-Tsung Tai, J. Phys. A {\bf 46}, 185301 (2013).
\bibitem{FSM3} S. Abe, Phys. Rev. A {\bf 46}, 1667 (1992).
\bibitem{FSM0} D. N. Page, Phys. Rev. A {\bf 36}, 3479 (1987).
\bibitem{FSM1} S. Kobayashi, K. Nomizu, {\it Fundations of Differential Geometry}, Vol. 2, Wiley, New York, (1969).
\bibitem{FSMqsgm} J. P. Provost, G. Valle, Commun. Math. Phys. {\bf 76}, 289 (1980).
\bibitem{FSMqsgm2} M. Revicule, M. Cassas, A. Plastino, Phys. Rev. {\bf 55}, 1695 (1997).
\bibitem{CSRS} D. C. Brody and E.-M. Graefe, J. Phys. A {\bf 43} 255205 (2010).
\bibitem{FSM4} M. Kolodrubetz, V. Gritsev and A. Polkovnikov, Phys. Rev. B {\bf 88}, 064304 (2013).
\bibitem{torus} A. R. Kuzmak, V. M. Tkachuk, J. Phys. A \textbf{49}, 045301 (2016).
\bibitem{FMM} A. R. Kuzmak, J. Geom. Phys. \textbf{116}, 81 (2017).
\bibitem{opticallattice1} L.-M. Duan, E. Demler, M. D. Lukin, Phys. Rev. Lett. {\bf 91}, 090402 (2003).
\bibitem{opticallattice18} A. B. Kuklov, B. V. Svistunov, Phys. Rev. Lett. {\bf 90}, 100401 (2003).
\bibitem{opticallattice17} J. Joo, Yuan Liang Lim, A. Beige, P. L. Knight, Phys. Rev. A {\bf 74}, 042344 (2006).
\bibitem{brach} A. R. Kuzmak, V. M. Tkachuk, J. Phys. A \textbf{46}, 155305 (2013).
\end{thebibliography}
\end{document}